\documentclass[aps,prl,reprint,superscriptaddress,longbibliography,nofootinbib]{revtex4-2}

\usepackage{amsmath,amssymb,mathtools,bm}
\usepackage{microtype}
\usepackage{hyperref}
\usepackage{xcolor}
\usepackage{booktabs}
\usepackage{enumitem}
\usepackage{physics}
\usepackage{url}
\usepackage{tikz}
\usetikzlibrary{arrows.meta,decorations.pathreplacing}

\hypersetup{colorlinks=true,linkcolor=blue!55!black,citecolor=blue!55!black,urlcolor=blue!55!black}
\allowdisplaybreaks
\newcommand{\Z}{\mathbb Z}
\newcommand{\ee}{\mathrm e}
\newcommand{\ii}{\mathrm i}
\newcommand{\diag}{\operatorname{diag}}
\newcommand{\AP}{\mathrm{AP}}

\newcommand{\half}{\tfrac12}

\begin{document}

\title{Microscopic Origin of the Shannon--R\'enyi Boundary Phase in the Critical Ising Chain}

\author{M. A. Rajabpour}
\affiliation{Instituto de F\'isica, Universidade Federal Fluminense, Av. Gal. Milton Tavares de Souza s/n, Gragoat\'a, 24210-346 Niter\'oi, RJ, Brazil}

\date{August 2026}

\begin{abstract}
Shannon--R\'enyi entropies probe the full measurement distribution of a quantum many-body state, but their microscopic evaluation remains difficult even in free-fermion systems. We show that the Born distribution of the critical transverse-field Ising chain is exactly the parity boundary of a weak inverse-square classical Ising model, a structure revealed by the Cauchy form of its probability kernel. The construction is exact at finite size for every real $n>0$ and isolates the long-distance sectors controlling the $n>1$ thermodynamics. Cutting a finite interval generates the universal $c\,n/[8(n-1)]$ logarithm directly from the missing $1/r^2$ interactions, with $c=1/2$ the Ising central charge. On the ring, the explicit finite-size factor has no constant term, while any additional noninteger singularity is confined to a residual cycle-only free energy. All explicit long-distance contributions become marginal only at $n=1$.
\end{abstract}

\maketitle

How can a basis-dependent distribution over exponentially many measurement outcomes encode universal boundary physics?  The Born probabilities of a many-body wave function are microscopic, yet their moments can retain universal information beyond the leading volume law.  This is well established in Shannon--R\'enyi and participation entropies and in multifractal properties of critical many-body wave functions \cite{Stephan2009,AtasBogomolny2012,AtasBogomolny2014,Luitz2014PRL}, with related configuration-space structures found across quantum spin chains and critical systems \cite{LuitzParticipation2014,Luitz2014PRB,Misguich2016,MisguichOshikawa2017,SierantTurkeshi2022}.  Such probability-level information is also becoming experimentally accessible through projective measurements and quantum snapshots \cite{Li2023Participation,Meurice2025,Wei2022}; notably, the central charge has been extracted from subleading Shannon--R\'enyi scaling on a universal quantum processor \cite{Koyluoglu2026}. Exact correspondences also connect these quantities to other resource measures such as stabilizer R\'enyi entropies \cite{Rajabpour2026PRL}.  More generally, averaging over measurement outcomes has emerged as a source of boundary, defect, and measurement-induced critical phenomena \cite{WeinsteinSajithAltmanGarratt2023,KhannaVasseur2026,PatilLudwig2024,KumarPatilLudwigVasseur2026,PatilLudwig2025,RajabpourSnapshots2026}.  A basic question is therefore whether the universal boundary structures seen at long distances can be identified directly in the microscopic Born distribution.

For Shannon--R\'enyi entropies this question is particularly sharp.  Universal subleading terms are closely tied to boundary free energies and R\'enyi-index-dependent boundary conditions \cite{Stephan2009,Stephan2010,Zaletel2011,Stephan2011}, with related boundary-CFT structures in conformal quantum critical wave functions \cite{FradkinMoore2006,Oshikawa2010}.  Periodic critical chains exhibit universal constants and R\'enyi-index transitions \cite{Stephan2011,Stephan2014,Misguich2016}, while subsystem Shannon and R\'enyi mutual informations display central-charge-controlled logarithms in Ising, Potts, Ashkin--Teller, and XXZ chains \cite{AlcarazRajabpour2013,AlcarazRajabpour2014,AlcarazRajabpour2015,Alcaraz2016}, with related basis-dependent finite-size structures in measured-spin and classical-Ising settings \cite{UmParkHinrichsen2012,LauGrassberger2013}.  Most directly, the marginal entropy of an interval of length $\ell$ in the infinite critical Ising chain contains the universal $c\,n/[8(n-1)]\log\ell$ term for $n>1$, and analogous logarithms occur in broad classes of critical quadratic fermions \cite{Stephan2014,AlcarazRajabpour2014,Tarighi2022}.  Replica boundary ordering gives the continuum interpretation of this branch, with related ordering phenomena rigorously established in critical book-Ising geometries \cite{Stephan2011,Stephan2014,DuminilCopinGarbanTassion2023}.  The universal coefficient and its boundary interpretation are therefore known.  What is not known is the microscopic mechanism by which the complete lattice probability distribution produces them.

That problem remains nontrivial even for free fermions.  Gaussianity makes individual configuration probabilities accessible through determinant, Toeplitz, and Pfaffian structures \cite{AbanovFranchini2003,FranchiniAbanov2005,StephanEFP2014,AresViti2020,NajafiRajabpour2016,TarighiKhassehRajabpour2024}, while related methods address full counting statistics, principal-minor sums, and leading Shannon--R\'enyi behavior \cite{Ivanov2013,Groha2018,NattaghNajafi2025,Monthus2015,Luitz2014PRB}.  But the R\'enyi moment is a nonlinear sum over exponentially many probabilities, and knowledge at integer replica index does not by itself determine the behavior at general real $n$ when genuine R\'enyi-index transitions are possible \cite{Stephan2011,Stephan2014}.  What is needed is a representation of the probability law itself that exposes the relevant long-distance structure before any thermodynamic or replica limit is taken.

\begin{figure}[t]
\centering
\begin{tikzpicture}[x=1cm,y=1cm,>=Latex,font=\sffamily\scriptsize]
  \begin{scope}[shift={(-2.15,0)}]
    \draw[line width=0.7pt] (0,0) circle (0.72);
    \foreach \a in {0,30,...,330}{\fill ({0.72*cos(\a)},{0.72*sin(\a)}) circle (1.35pt);}
    \node at (0,-1.02) {(a) full periodic chain};
    \node at (0,0) {$L$};
    \draw[->,thin] (0.82,0.20) arc[start angle=14,end angle=-36,radius=0.82];
  \end{scope}
  \begin{scope}[shift={(2.05,0)}]
    \draw[line width=0.65pt] (-1.75,0)--(1.75,0);
    \foreach \x in {-1.55,-1.25,-0.95,-0.65,-0.35,-0.05,0.25,0.55,0.85,1.15,1.45}{\fill (\x,0) circle (1.35pt);}
    \draw[blue!65!black,line width=1.5pt] (-0.78,0)--(0.98,0);
    \foreach \x in {-0.65,-0.35,-0.05,0.25,0.55,0.85}{\fill[blue!65!black] (\x,0) circle (1.55pt);}
    \draw[red!70!black,densely dashed] (-0.80,-0.35)--(-0.80,0.42);
    \draw[red!70!black,densely dashed] (1.00,-0.35)--(1.00,0.42);
    \draw[decorate,decoration={brace,amplitude=3.5pt,mirror}] (-0.78,-0.30)--(0.98,-0.30)
      node[midway,below=4pt] {$A_\ell$, length $\ell$};
    \node at (0.10,-1.02) {(b) interval of an infinite chain};
    \node at (-1.88,0) {$\cdots$}; \node at (1.88,0) {$\cdots$};
  \end{scope}
\end{tikzpicture}
\caption{The two geometries are evaluated from the same critical Born distribution.  Periodicity removes a surface, whereas the two cuts of $A_\ell$ expose the long-range tail of the auxiliary interaction.}
\label{fig:geometries}
\end{figure}
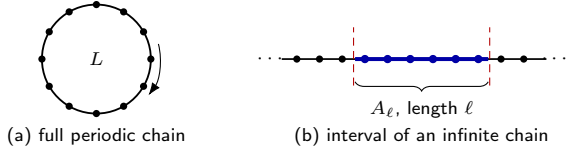

Here we find such a representation for the critical transverse-field Ising chain.  A Cauchy structure hidden in its Gaussian probability kernel turns the full family of principal minors into Boltzmann weights of a weak ferromagnetic inverse-square Ising model.  More strikingly, the Born distribution itself is exactly the parity boundary of an independent long-range edge ensemble: the vertices of odd occupied degree occur with precisely the quantum measurement probabilities.  This identity is exact at finite size and does not rely on replicas.  It makes the relation between geometry and entropy transparent.  The periodic chain and a finite interval are two realizations of the same microscopic measure, but cutting the interval exposes the inverse-square tail and generates the known universal logarithm directly on the lattice.  The same construction extends to arbitrary real $n>0$ at finite size and isolates the remaining nontrivial thermodynamics into a well-defined residual sector.  Universal boundary physics is thus traced to a concrete structure already present in the microscopic Born probabilities.  We consider the critical TFI chain

\begin{equation}
 \mathcal H(h)=-\frac12\sum_{j=0}^{L-1}\left(\sigma_j^x\sigma_{j+1}^x+h\sigma_j^z\right),
 \qquad h=1,
 \label{eq:Hmain}
\end{equation}
with periodic boundary conditions, even $L$, and in the $\sigma^z$ basis.  For a measured region $A$, let $p_A(\bm\sigma_A)$ be its marginal Born distribution and define
\begin{equation}
 H_n(A)=\frac{1}{1-n}\log\sum_{\bm\sigma_A}p_A(\bm\sigma_A)^n,
 \qquad n>0.
 \label{eq:defHn}
\end{equation}
We treat the two geometries in Fig.~\ref{fig:geometries}: the complete periodic chain and a finite interval $A_\ell$ cut from the infinite critical chain.

The lattice construction has three layers.  First, it is exact at finite size: every principal minor of the critical Ising probability kernel becomes one Boltzmann weight of a weak ferromagnetic inverse-square Ising model.  Second, for every integer $n\ge2$ the full moment is an exact $\mathbb Z_2^{n-1}$ current partition function, while the single-edge sector can be evaluated in closed form.  Third, the underlying probability law admits a replica-free representation for every real $n>0$.  This last identity permits a controlled continuation analysis without analytically continuing the replica group away from integer $n$.

We now derive this structure directly from the Born probabilities.

\paragraph{Cauchy minors and a hidden Ising model.---}
For Eq.~\eqref{eq:Hmain}, the critical ground-state probabilities in the $\sigma^z$ basis can be written \cite{Pfeuty1970,Stephan2010}
\begin{equation}
 p_{\bm\sigma}=2^{-L}\det(I+D_{\bm\sigma}G),\quad
 G_{j\ell}=\frac{(-1)^{j-\ell}}{L\sin[\pi(j-\ell+\half)/L]} ,
 \label{eq:pdet}
\end{equation}
where $D_{\bm\sigma}=\diag(\sigma_0^z,\ldots,\sigma_{L-1}^z)$.  Expanding the determinant in principal minors gives the Walsh representation
\begin{equation*}
 p_{\bm\sigma}=2^{-L}\sum_{S\subseteq[L]}d_L(S)\chi_S(\bm\sigma),
 \qquad
 d_L(S)=\det G[S,S],
\end{equation*}
where $\chi_S(\bm\sigma)\equiv\prod_{j\in S}\sigma_j^z$ is the Walsh character associated with the subset $S$.  Thus the complete Born distribution is encoded in the family of principal minors $d_L(S)$.  A diagonal gauge removes the alternating signs without changing these principal minors.  The resulting matrix has Cauchy form, and the trigonometric Cauchy identity \cite{Krattenthaler1999} gives
\begin{equation}
 \begin{gathered}
 d_L(S)=A_L^{|S|}\prod_{i<j\in S}g_{j-i}^{(L)},\qquad
 A_L=\frac{1}{L\sin(\pi/2L)},\\
 g_m^{(L)}=\frac{\sin^2(\pi m/L)}
 {\sin^2(\pi m/L)-\sin^2(\pi/2L)} .
 \end{gathered}
 \label{eq:cauchyproduct}
\end{equation}
The elementary sine products imply
$A_L^2\prod_{m=1}^{L-1}g_m^{(L)}=1$.
Writing $\tau_j=(-1)^{\mathbf1(j\in S)}$ therefore cancels every one-spin term in $\log d_L(S)$ and gives the exact identity
\begin{equation}
 d_L(S)=A_L^{L/4}
 \exp\!\left[\sum_{i<j}J_{j-i}^{(L)}\tau_i\tau_j\right],\quad
 J_m^{(L)}=\frac14\log g_m^{(L)} .
 \label{eq:oneminorising}
\end{equation}

Equation~\eqref{eq:oneminorising} has a direct statistical-mechanical interpretation.  The Walsh expansion of the Born distribution contains one principal minor for every subset $S$; the Cauchy factorization turns the full many-body dependence of each coefficient into pairwise interactions, while the product identity cancels its one-body field exactly on the ring.  An exponentially large family of Gaussian minors is therefore generated by one translation-invariant classical interaction.  This is the microscopic mechanism behind both the replica construction and its replica-free form.

This identity is the starting point for the constructions below.  In the thermodynamic limit $J_m=\frac14\log[4m^2/(4m^2-1)]$.  The generated interaction lies in the classical inverse-square Ising family \cite{Dyson1969,FrohlichSpencer1982,Imbrie1982}, but is weak: its high-temperature norm, $\rho_J\equiv2\sum_{m\ge1}\tanh J_m=0.2255407656\ldots<1$, places the auxiliary model well inside a standard uniqueness regime \cite{Dobrushin1968,Ruelle1969,Simon1993,KoteckyPreiss1986,Ueltschi2004}.

\paragraph{The ring: integer replicas and the periodic constant.---}
Expanding Eq.~\eqref{eq:pdet} in Walsh characters and summing the $n$th power for integer $n\ge2$ imposes one local XOR constraint.  Equivalently, each site carries an even-parity vector $\bm\tau_j\in G_n\simeq\Z_2^{n-1}$ and the replica partition function has pair energy $J_m\sum_{a=1}^n\tau_i^{(a)}\tau_j^{(a)}$.  Fourier transforming the finite Abelian group gives a conserved colored-current gas.  Extracting its trivial character yields
\begin{equation}
 M_n(L)\equiv\sum_{\bm\sigma}p_{\bm\sigma}^{\,n}
 =\left[\prod_{m=1}^{L-1}B_n(z_m^{(L)})\right]^{L/2}\Xi_{n,L},
 \label{eq:Bninteger}
\end{equation}
where
\begin{align}
 B_n(z)&=2^{-n}\!\left[(1+\sqrt{1-z})^n
 +(1-\sqrt{1-z})^n\right],\nonumber\\
 z_m^{(L)}&=\frac{\sin^2(\pi/2L)}{\sin^2(\pi m/L)} .
 \label{eq:Bn}
\end{align}

The character transform is useful because it cleanly separates two kinds of physics.  $B_n$ contains everything that can be assigned independently to one separation $m$, whereas $\Xi_{n,L}$ begins with genuinely connected current loops.  In the special case $n=2$, this decomposition is the familiar high-temperature even-subgraph expansion; for $n=3$, the local group has four elements and the auxiliary model is exactly a ferromagnetic four-state long-range Potts model (see Supplemental Material~\cite{SM}).  Thus the construction is not a formal replica bookkeeping device: at each integer it produces a conventional classical statistical model with explicit positive couplings.

For integer $n$, $B_n$ factorizes into $\lfloor n/2\rfloor$ linear factors,
$B_n(z)=\prod_r(1-\lambda_{n,r}z)$ with
$\lambda_{n,r}=\cos^2[(2r-1)\pi/(2n)]$.

The factorization also turns the finite-size problem into an elementary product.  If $\sin\phi_{r,L}=\sqrt{\lambda_{n,r}}\sin(\pi/2L)$, then
\begin{equation}
 \prod_{m=1}^{L-1}(1-\lambda_{n,r}z_m^{(L)})
 =\left[\frac{\sin(L\phi_{r,L})}{L\sin\phi_{r,L}}\right]^2 .
 \label{eq:finiteprodmain}
\end{equation}
Because $L\phi_{r,L}=\pi a_{n,r}+O(L^{-2})$, the logarithm of this factor is $L$ times a finite limit plus $O(L^{-1})$.  In particular, there is no room for a constant hidden in an Euler--Maclaurin remainder.

A finite sine-product identity therefore evaluates the first factor in Eq.~\eqref{eq:Bninteger} exactly: its contribution to the entropy is $H_{n,0}^{\rm PBC}(L)=Ls_{n,0}+O(L^{-1})$, with no $L^0$ term.  The remaining factor $\Xi_{n,L}$ is a connected periodic current problem.  For every fixed integer $n$ its bulk pressure exists because the underlying pair interaction is absolutely summable; at each fixed current order, periodicity gives exactly $L$ anchors and no surface term, while winding contributions vanish with the inverse-square tail.  At $n=2$ the high-temperature norm is small and the linked-cluster coefficients decrease rapidly; the fixed-order expansion has no periodic surface term and agrees with the established vanishing intercept.  A fully rigorous promotion of this statement to the particular all-orders Mayer series requires an explicit polymer-convergence criterion, so we do not infer that criterion from the Dobrushin bound alone.  For general $n$, summing the fixed-order statement uniformly to all orders is likewise a separate thermodynamic question.  The explicit part of the density is
\begin{align}
 s_{n,0}&=\frac{1}{n-1}\sum_{r=1}^{\lfloor n/2\rfloor}
 \log\!\frac{\pi a_{n,r}}{\sin(\pi a_{n,r})},\nonumber\\
 a_{n,r}&=\frac12\cos\frac{(2r-1)\pi}{2n}.
 \label{eq:intdensity}
\end{align}
where $\psi_n\equiv\lim_{L\to\infty}L^{-1}\log\Xi_{n,L}$, so $s_n=s_{n,0}-\psi_n/(n-1)$.  For $n=2$ the current is the even-subgraph gas with $\tanh K_m=(8m^2-1)^{-1}$; its rapidly convergent linked-cluster series reproduces $s_2=0.2138074203\ldots$ directly on the lattice \cite{Stephan2010}.

Here $\psi_n$ is the thermodynamic pressure of the residual connected colored-current sector, not a fitted correction to an otherwise exact answer.  At $n=2$ the first few Eulerian clusters already saturate the density to high precision; at general integer $n$ the character expansion supplies the analogous fixed-order algorithm (see Supplemental Material~\cite{SM}).  Thus the same microscopic construction that exposes the universal boundary structure also gives an exact lattice representation of the nonuniversal volume coefficient as an explicit contribution plus a residual pressure.  Related Macdonald--Ruijsenaars and Selberg formulations are given in Supplemental Material \cite{Macdonald1995,Ruijsenaars1987,Selberg1944,deBruijn1955,ForresterWarnaar2008,SM}.

\paragraph{Replica-free continuation to real $n$.---}
Continuing $G_n$ or the finite product in Eq.~\eqref{eq:intdensity} would be unjustified.  Equation~\eqref{eq:oneminorising} avoids that problem.  If $\mathcal D_{\bm\sigma}=\{j:\sigma_j=-1\}$ and
\begin{equation}
 W_L(D)=\sum_{\{\tau\}}\ee^{\sum_{i<j}J_{j-i}^{(L)}\tau_i\tau_j}
 \prod_{j\in D}\tau_j,
 \end{equation}
then Walsh inversion gives the \emph{probability itself},
$p_{\bm\sigma}=2^{-L}A_L^{L/4}W_L(\mathcal D_{\bm\sigma})$.
Consequently
\begin{equation}
 M_n(L)=2^{-nL}A_L^{nL/4}\sum_{D\subseteq[L]}W_L(D)^n
 \label{eq:realmoment}
\end{equation}
for every real $n>0$; odd $D$ vanish and ferromagnetic positivity makes the remaining powers unambiguous.  
Equivalently, writing $p_D\equiv p_{\bm\sigma}$ for $D=\mathcal D_{\bm\sigma}$ and expanding the auxiliary Ising model in $t_e=\tanh J_e^{(L)}$ gives
\begin{equation}
 p_D=\frac{\displaystyle\sum_{F:\,\partial F=D}\prod_{e\in F}t_e}
 {\displaystyle\prod_e(1+t_e)} .
 \label{eq:paritymain}
\end{equation}
Hence one may occupy every auxiliary edge independently with probability $q_e=t_e/(1+t_e)$ and retain only the set $D=\partial F$ of vertices with odd occupied degree.  That parity boundary is \emph{exactly} the Ising Born distribution.  This representation proves positivity without referring to a wave-function sign convention and shows that the real power in Eq.~\eqref{eq:realmoment} is an ordinary power of a classical probability law.

Thus the continuation is not inferred from integer replicas: Eq.~\eqref{eq:realmoment} is the physical R\'enyi moment itself for every finite $L$.  This distinction is essential near $n=1$.  Continuing the integer polynomial $B_n$ or the group $G_n$ would say nothing about a thermodynamic singularity between replica points; the real-power sum instead starts from the original positive Born measure before any limit is taken.  It converts the analytic-continuation question into a concrete statistical-mechanics question: after all acyclic contributions have been removed, does the residual cyclic free energy remain on one branch?

The same normalization $B_n$ in Eq.~\eqref{eq:Bn} is now an analytic function of real $n$.  Its infinite-volume contribution is the convergent series
\begin{equation}
 s_{n,0}=-\frac{1}{n-1}\sum_{m=1}^{\infty}
 \log B_n\!\left(\frac{1}{4m^2}\right),\qquad n>1,
 \label{eq:realdensity}
\end{equation}
which reduces exactly to Eq.~\eqref{eq:intdensity} at integer $n$.  
The small auxiliary coupling rigorously controls the underlying Ising correlations and is indispensable for the exact $n=2$ expansion.  For noninteger powers, however, smallness of the auxiliary high-temperature norm is not by itself a polymer theorem for the real-power source sum.  The independent-edge representation gives the sharper separation needed below: it proves exact factorization on every forest and confines the genuinely new thermodynamics to cycle space.

The independent-edge form also identifies exactly what remains after the one-edge factor is removed.  On a forest the boundary map from occupied edges to odd-degree vertices is injective, so the full real-$n$ moment factorizes into its one-edge moments for every $n>0$; the residual factor is generated only by cyclic cores (see Supplemental Material~\cite{SM}).  This finite-graph statement is exact and replaces a formal continuation of the replica current gas.  For each fixed cyclic-core order, the inverse-square tail makes the periodic correction summable for $n>1$.  What it does not by itself provide is a uniform all-orders polymer theorem arbitrarily close to $n=1$.  We therefore distinguish the exact replica-free continuation and its fixed-order control from the additional nonperturbative assumption that the residual cyclic pressure stays on one analytic branch throughout $n>1$.  Existing boundary-phase results and direct finite-size checks support that continuation.  Direct finite-size diagnostics of the residual factor for noninteger $1<n<2$ provide additional support (see Supplemental Material~\cite{SM}).

The explicit real-index factor also identifies the endpoint.  For noninteger $n$,
\begin{equation}
 B_n(z)=1-\frac n4z+\cdots+4^{-n}z^n[1+O(z)] .
 \label{eq:nonanalyticBn}
\end{equation}

This branch is the point at which the explicit real-index factor contains information that cannot be obtained by inspecting integers.  Near a ring endpoint $z_m^{(L)}\sim(2m)^{-2}$, so the $z^n$ term probes the sum of a power-law tail $m^{-2n}$.  Comparing the finite chord sum with the two infinite-line endpoints gives the scale $L\sum_{m\gtrsim L}m^{-2n}\sim L^{2-2n}$.

On the ring this branch term gives an $L^{2-2n}$ correction after the bulk piece is removed.  It dominates the ordinary $1/L$ correction for $1<n<3/2$, meets it at $n=3/2$, and is subleading above.  Only at $n=1$ does this explicit branch become $L^0$.  Hence $n=3/2$ is only a correction-to-scaling crossover of the explicit factor; any additional transition would have to come from the residual cyclic free energy.

\paragraph{A finite interval: the logarithm from missing $1/r^2$ pairs.---}
Take now the infinite critical chain and retain $A_\ell=\{0,\ldots,\ell-1\}$.  The marginal probability again has the determinant form, with the compressed Cauchy kernel $O_{ij}=[\pi(i-j+\half)]^{-1}$.  Its minors obey Eq.~\eqref{eq:cauchyproduct} with
$A=2/\pi$ and $g_m=4m^2/(4m^2-1)$.  The Wallis identity $A\prod_{m\ge1}g_m=1$ converts the compression into the same long-range Ising model, now with two explicit boundary fields
\begin{align}
 h_j^{(\ell)}&=\frac12\!\left(\sum_{m=j+1}^{\infty}K_m+
 \sum_{m=\ell-j}^{\infty}K_m\right),\nonumber\\
 K_m&=\frac12\log\frac{4m^2}{4m^2-1} .
 \label{eq:blockfield}
\end{align}
For a single boundary
$h_d=\frac14\log[\Gamma(d+\half)\Gamma(d+\frac32)/\Gamma(d+1)^2]
\sim(16d)^{-1}$ as $d\to\infty$.

The sign and spatial profile of Eq.~\eqref{eq:blockfield} are also instructive.  Deep inside a long block both missing tails vanish and the field tends to zero, recovering the translation-invariant bulk auxiliary model; at fixed distance from a cut a positive field survives.  Thus compression does not create an unknown boundary condition: it creates a calculable one.  The long $h_d\sim1/d$ tail looks potentially dangerous, but its potentially nonanalytic real-$n$ contribution is proportional to $h_d^n$ and is summable exactly when $n>1$.

At integer $n$, the same group-character expansion separates a bulk pair factor, explicit boundary factors, and a connected boundary current pressure.  The replica-free parity representation gives the same one-edge factors directly for every real $n>0$; hence for real $n>1$ the explicit pair contribution is governed by
\begin{align}
 f_n(m)&=\frac n2K_m+\log\!\left[\cosh^n\!\frac{K_m}{2}
 +\sinh^n\!\frac{K_m}{2}\right],\nonumber\\
 &=\frac{n}{16m^2}+O(m^{-\min(4,2n)}).
 \label{eq:fntail}
\end{align}

Equation~\eqref{eq:fntail} gives a direct microscopic derivation of the universal coefficient.  The leading $1/m^2$ term is analytic in $n$ and fixed solely by the half-shift Cauchy geometry.  All noninteger structure begins at order $m^{-2n}$; after multiplication by the number $m$ of pairs lost to a cut, its sum converges for every $n>1$.  Therefore there is exactly one marginal harmonic series, and its coefficient is already determined before any connected-loop calculation is performed.

The interval has only $\ell-m$ pairs at separation $m$.  Relative to the bulk pressure, the missing-pair contribution is therefore
\begin{equation}
 \sum_{m<\ell}m f_n(m)+\ell\sum_{m\ge\ell}f_n(m)
 =\frac n{16}\log\ell+C_{\rm pf}(n)+o(1),
 \label{eq:missingpairs}
\end{equation}
where $C_{\rm pf}(n)$ is the convergent pair constant given explicitly in Supplemental Material~\cite{SM}.  Since this missing-pair term enters the entropy with the opposite sign to the bulk pair pressure, its logarithmic contribution is
\begin{equation*}
 \frac{n}{16(n-1)}\log\ell
 =\frac{c\,n}{8(n-1)}\log\ell,\qquad c=\frac12 .
\end{equation*}
This proves the logarithm generated by the explicit pair sector without CFT and reproduces the established Ising coefficient \cite{Stephan2014,AlcarazRajabpour2014,Tarighi2022}.  The remainder of this sector is summable precisely for $n>1$.  The boundary field provides the same test: its nonanalytic contribution behaves as $\sum_d h_d^n\sim\sum_d d^{-n}$ and hence becomes marginal only at $n=1$.

The same separation shows why the two geometries must share the same bulk pressure whenever the real-index cyclic pressure exists: clusters anchored a macroscopic distance from either cut see the same translation-invariant environment as on the ring.  What changes is the surface sector.  The interval therefore acquires a logarithm and a boundary constant without changing the thermodynamic density.

The interval $O(1)$ term admits the same separation: an explicit, absolutely convergent pair constant and boundary factor, plus a residual cycle-only boundary excess.  The complete formula, its Barnes-$G$ reduction at integer $n$, and the precise replica-free definition of the residual excess are given in Supplemental Material~\cite{SM}.  The associated Lee--Yang polynomial is also given there \cite{LeeYang1952,YangLee1952,SM}.  Thus the logarithm and all potentially marginal explicit tails are fixed before the only remaining all-orders continuation question enters.

\paragraph{Discussion.---}
A basis-resolved Born distribution can therefore carry the full long-distance structure required for universal boundary physics.  In the critical Ising chain this structure is exposed by an exact classical representation: the same parity-boundary measure describes both the ring and the interval, and changing the geometry simply reorganizes its long-range interactions.  The boundary logarithm is then not inserted through a continuum boundary condition; it emerges from the microscopic probability law itself.

The exact finite-size representation holds for every real $n>0$, while the thermodynamic decomposition developed here is controlled for $n>1$.  The explicit one-edge and boundary sectors become marginal only at $n=1$; any additional singularity for $n>1$ would have to arise nonperturbatively from the residual cycle sector.  Establishing uniform all-orders control of that sector, particularly near the Shannon point, remains an open problem.

More broadly, the present construction suggests that universal information in measurement distributions may be understood by first identifying the effective graphical or statistical-mechanical structure hidden in their microscopic probabilities.  In the Ising case, an inverse-square interaction converts a spatial cut into a harmonic boundary cost and hence into the universal logarithm.  Continuum boundary data can therefore be encoded, and read directly, in the microscopic organization of a quantum Born distribution.

{\it Acknowledgements:}
M.A.R. acknowledges partial support from CNPq and FAPERJ (grant number E-26/210.062/2023).

\clearpage
\onecolumngrid
\setcounter{equation}{0}
\renewcommand{\theequation}{S\arabic{equation}}
\setcounter{figure}{0}
\renewcommand{\thefigure}{S\arabic{figure}}
\setcounter{table}{0}
\renewcommand{\thetable}{S\arabic{table}}
\setcounter{section}{0}
\renewcommand{\thesection}{S\arabic{section}}
\renewcommand{\thesubsection}{\thesection.\arabic{subsection}}

\begin{center}
{\Large\bf Supplemental Material for\\[2mm]
``Microscopic Origin of the Shannon--R\'enyi Boundary Phase in the Critical Ising Chain''}\\[3mm]
M. A. Rajabpour$^{2}$\\[1mm]

$^{2}$Instituto de F\'isica, Universidade Federal Fluminense, Av. Gal. Milton Tavares de Souza s/n, Gragoat\'a, 24210-346 Niter\'oi, RJ, Brazil
\end{center}

\bigskip

This Supplemental Material gives the complete lattice derivation used in the Letter and records additional structures that are useful beyond the minimal PRL narrative.  No conformal-field-theory assumption is used in obtaining the asymptotic coefficients.  BCFT results are used only for comparison after the lattice calculation.  Site labels on a ring are understood modulo $L$ and $L$ is even.

\section{Critical TFI probabilities and the Walsh transform}

We use
\begin{equation}
\mathcal H(h)=-\frac12\sum_{j=0}^{L-1}\left(\sigma_j^x\sigma_{j+1}^x+h\sigma_j^z\right),
\qquad \sigma_L^a\equiv\sigma_0^a,
\label{eqS:H}
\end{equation}
and set $h=1$.  The critical TFI ground state is Gaussian \cite{Pfeuty1970}.  In the computational ($\sigma^z$) basis the diagonal probabilities admit the determinant representation \cite{Stephan2010}
\begin{equation}
p_{\bm\sigma}=2^{-L}\det(I+D_{\bm\sigma}G_L),
\qquad
(G_L)_{j\ell}=\frac{(-1)^{j-\ell}}{L\sin[\pi(j-\ell+\half)/L]},
\label{eqS:pdet}
\end{equation}
where $D_{\bm\sigma}=\diag(\sigma_0^z,\ldots,\sigma_{L-1}^z)$.
For any matrix $G$,
\begin{equation}
\det(I+D_{\bm\sigma}G)
=\sum_{S\subseteq[L]}\chi_S(\bm\sigma)\det G[S,S],
\qquad
\chi_S(\bm\sigma)=\prod_{j\in S}\sigma_j^z .
\label{eqS:Walsh}
\end{equation}
The $\chi_S$ are the Walsh characters of the Boolean cube and obey
\begin{equation}
2^{-L}\sum_{\bm\sigma}\chi_S(\bm\sigma)\chi_T(\bm\sigma)=\delta_{S,T}.
\label{eqS:WalshOrth}
\end{equation}
For $n=2$ this immediately gives Parseval,
\begin{equation}
M_2(L)=\sum_{\bm\sigma}p_{\bm\sigma}^2
=2^{-L}\sum_{S\subseteq[L]}\det G_L[S,S]^2 .
\label{eqS:parseval}
\end{equation}
For an integer $n>2$, multiplying $n$ Walsh expansions instead gives the XOR condition
\begin{equation}
M_n(L)=2^{-(n-1)L}
\sum_{\substack{S_1,\ldots,S_n\subseteq[L]\\
S_1\triangle\cdots\triangle S_n=\varnothing}}
\prod_{a=1}^n d_L(S_a),
\qquad d_L(S)=\det G_L[S,S].
\label{eqS:XOR}
\end{equation}
Equation~\eqref{eqS:XOR} is the exact starting point for integer replicas.

For a block $A_\ell\subset[L]$, summing Eq.~\eqref{eqS:Walsh} over every spin outside $A_\ell$ kills all characters whose support intersects the complement.  Hence the marginal distribution is obtained by simply compressing the matrix:
\begin{equation}
p_{A_\ell}(\bm\sigma_A)=2^{-\ell}
\det\!\left(I_\ell+D_{\bm\sigma_A}G_L[A_\ell,A_\ell]\right).
\label{eqS:blockfiniteL}
\end{equation}
Taking $L\to\infty$ at fixed $\ell$ therefore gives the exact interval formula used below; no assumption about a reduced Gaussian wave function is needed.

\section{Trigonometric Cauchy factorization}

Introduce $D_\pi=\diag(1,-1,1,-1,\ldots)$ and $O_L=D_\pi G_LD_\pi$.  Principal minors are unchanged, while
\begin{equation}
(O_L)_{j\ell}=\frac{1}{L\sin(a_j-a_\ell+\delta)},
\qquad a_j=\frac{\pi j}{L},\quad \delta=\frac{\pi}{2L}.
\label{eqS:O}
\end{equation}
The standard trigonometric Cauchy determinant \cite{Krattenthaler1999}
\begin{equation}
\det_{p,q}\frac1{\sin(x_p-y_q)}=
\frac{\prod_{p<q}\sin(x_p-x_q)\sin(y_q-y_p)}
{\prod_{p,q}\sin(x_p-y_q)}
\label{eqS:trigCauchy}
\end{equation}
(up to the overall sign fixed by the one-dimensional case) gives, for any $S\subseteq[L]$,
\begin{equation}
d_L(S)=A_L^{|S|}\prod_{\substack{i<j\\i,j\in S}}g_{j-i}^{(L)},
\qquad
A_L=\frac{1}{L\sin\delta},
\label{eqS:minorprod}
\end{equation}
where
\begin{equation}
g_m^{(L)}=
\frac{\sin^2(\pi m/L)}
{\sin[\pi(m-\half)/L]\sin[\pi(m+\half)/L]}
=\frac{\sin^2(\pi m/L)}{\sin^2(\pi m/L)-\sin^2\delta}>1 .
\label{eqS:gL}
\end{equation}
The two elementary products
\begin{equation}
\prod_{m=1}^{L-1}\sin\frac{\pi m}{L}=\frac{L}{2^{L-1}},
\qquad
\prod_{m=0}^{L-1}\sin\frac{\pi(m+\half)}{L}=2^{1-L},
\label{eqS:sineproducts}
\end{equation}
imply the crucial identity
\begin{equation}
A_L^2\prod_{m=1}^{L-1}g_m^{(L)}=1.
\label{eqS:productidentity}
\end{equation}
All later zero-field cancellations are consequences of Eq.~\eqref{eqS:productidentity}.

The infinite-line limit is particularly simple:
\begin{equation}
O_{ij}=\frac{1}{\pi(i-j+\half)},\qquad
A=\frac2\pi,\qquad
g_m=\frac{4m^2}{4m^2-1},
\label{eqS:lineCauchy}
\end{equation}
with Wallis' product
\begin{equation}
A\prod_{m=1}^{\infty}g_m=1.
\label{eqS:Wallis}
\end{equation}

\section{Root-of-unity form and Macdonald--Ruijsenaars coefficient}

The same finite Cauchy matrix has a useful root-of-unity form that makes contact with discrete Selberg sums and difference operators.  Introduce
\begin{equation}
x_j=e^{2\pi i j/L},\qquad \tau=e^{-i\pi/L},
\label{eqS:rootvars}
\end{equation}
and
\begin{equation}
C_{ij}=\frac{(1-\tau)x_i}{x_i-\tau x_j}.
\label{eqS:Croot}
\end{equation}
A direct Fourier/Cauchy calculation gives
\begin{equation}
CC^\dagger=\gamma_L I,
\qquad
\gamma_L=L^2\sin^2\frac{\pi}{2L},
\label{eqS:Cunitary}
\end{equation}
so $U=C/\sqrt{\gamma_L}$ is unitary and diagonally gauge-equivalent to $O_L$.

For a subset $S$ of size $k$, define the root-of-unity cross-ratio
\begin{equation}
V_S(\tau)=
\prod_{\substack{i\in S\\j\notin S}}
\frac{x_i-\tau x_j}{x_i-x_j}.
\label{eqS:VS}
\end{equation}
The Cauchy determinant identity gives
\begin{equation}
\det C[S,S]
=\left[\frac{L(1-\tau)}2\right]^k
\tau^{\binom{k}{2}}V_S(\tau),
\label{eqS:CV}
\end{equation}
and all prefactors cancel after unitary normalization:
\begin{equation}
|V_S(\tau)|^2=|\det U[S,S]|^2=\det O_L[S,S]^2.
\label{eqS:Vminor}
\end{equation}
This is the exact bridge between the root-of-unity/Selberg-looking formulation and the real orthogonal Cauchy formulation used in the thermodynamic analysis.

There is also a direct relation to the coefficient of the $k$th Macdonald--Ruijsenaars difference operator \cite{Macdonald1995,Ruijsenaars1987}.  With
\begin{equation}
A_S(x;t)=t^{\binom{k}{2}}
\prod_{\substack{i\in S\\j\notin S}}
\frac{t x_i-x_j}{x_i-x_j},
\label{eqS:ASMac}
\end{equation}
one has exactly, at $t=\tau^{-1}$,
\begin{equation}
V_S(\tau)=
\tau^{k(L-k)+\binom{k}{2}}A_S(x;\tau^{-1}).
\label{eqS:MacdonaldRelation}
\end{equation}
The observation is structural rather than a solution of the entropy sum: standard Macdonald orthogonality sums independent row and column subsets, whereas the Shannon--R\'enyi problem retains the diagonal constraint $S=T$.  This explains both the usefulness and the limitation of the connection.

\section{Every principal minor is one Ising Boltzmann weight}

Let $x_j=\mathbf1(j\in S)$ and $\tau_j=(-1)^{x_j}$, so $x_j=(1-\tau_j)/2$.  Define
\begin{equation}
K_m^{(L)}=\frac12\log g_m^{(L)},
\qquad
J_m^{(L)}=\frac12K_m^{(L)}=\frac14\log g_m^{(L)}.
\label{eqS:KJ}
\end{equation}
Taking the logarithm of Eq.~\eqref{eqS:minorprod},
\begin{align}
\log d_L(S)
&=\log A_L\sum_jx_j+2\sum_{i<j}K_{j-i}^{(L)}x_ix_j\\
&=C_L+\sum_{i<j}J_{j-i}^{(L)}\tau_i\tau_j
-\frac12\sum_j\tau_j\left[\log A_L+\sum_{m=1}^{L-1}K_m^{(L)}\right].
\label{eqS:minoralgebra}
\end{align}
Equation~\eqref{eqS:productidentity} says that the square bracket vanishes.  The constant is
$C_L=(L/4)\log A_L$.  Therefore
\begin{equation}
d_L(S)=A_L^{L/4}
\exp\!\left[\sum_{i<j}J_{j-i}^{(L)}\tau_i\tau_j\right].
\label{eqS:singleIsing}
\end{equation}
The exactness of Eq.~\eqref{eqS:singleIsing} is important: the auxiliary Ising model is not introduced after replicas; it is already present in one Walsh coefficient.

In the thermodynamic limit
\begin{equation}
J_m=\frac14\log\frac{4m^2}{4m^2-1}
=\frac{1}{16m^2}+O(m^{-4}).
\label{eqS:Jline}
\end{equation}
Its high-temperature norm is
\begin{equation}
\rho_J\equiv2\sum_{m=1}^{\infty}\tanh J_m
=0.225540765579077\ldots<1.
\label{eqS:rhoJ}
\end{equation}
Inverse-square Ising interactions have a classic and subtle phase-transition theory \cite{Dyson1969,FrohlichSpencer1982,Imbrie1982}; the coupling generated here lies on the weak-coupling side.  Equation~\eqref{eqS:rhoJ} is a quantitative high-temperature control parameter for the auxiliary Ising correlations; by itself it is not an all-orders convergence theorem for the noninteger source sum.

\subsection{Exact random-current/parity representation of the Born distribution}

Let $\mathcal D_{\bm\sigma}=\{j:\sigma_j^z=-1\}$.  Since
$\chi_S(\bm\sigma)=\prod_{j\in \mathcal D_{\bm\sigma}}\tau_j$, inserting Eq.~\eqref{eqS:singleIsing} into Eq.~\eqref{eqS:Walsh} yields
\begin{equation}
p_D=2^{-L}A_L^{L/4}W_L(D),
\qquad
W_L(D)=\sum_{\{\tau\}}\ee^{\sum_{i<j}J_{j-i}^{(L)}\tau_i\tau_j}
\prod_{j\in D}\tau_j .
\label{eqS:pW}
\end{equation}
Thus $W_L(D)=Z_J\langle\prod_{j\in D}\tau_j\rangle_J$ is an unnormalized correlation function of one ferromagnetic Ising model.  Spin-flip symmetry gives $p_D=0$ for odd $|D|$.

There is an even more elementary representation.  Put $t_e=\tanh J_e$.  The high-temperature identity
$\ee^{J_e\tau_i\tau_j}=\cosh J_e(1+t_e\tau_i\tau_j)$ gives
\begin{equation}
W_L(D)=2^L\prod_e\cosh J_e
\sum_{F:\,\partial F=D}\prod_{e\in F}t_e,
\label{eqS:currentW}
\end{equation}
where $\partial F$ is the set of odd-degree vertices of the selected edge set $F$.  The product identity also implies
\begin{equation}
A_L^{L/4}\prod_e\cosh J_e
=\prod_e(1+t_e)^{-1}.
\label{eqS:normcurrent}
\end{equation}
Consequently
\begin{equation}
p_D=\frac{\displaystyle\sum_{F:\,\partial F=D}\prod_{e\in F}t_e}
{\displaystyle\prod_e(1+t_e)}.
\label{eqS:paritylaw}
\end{equation}
Equivalently, occupy each edge independently with probability
\begin{equation}
q_e=\frac{t_e}{1+t_e},
\label{eqS:qedge}
\end{equation}
and record the vertices of odd occupied degree.  Their parity boundary is distributed \emph{exactly} according to the critical TFI Born probabilities.  This representation simultaneously proves positivity and explains the global even-parity support.

\section{Integer \texorpdfstring{$n$}{n}: the \texorpdfstring{$\mathbb Z_2^{n-1}$}{Z2 replica-current} model}

Insert Eq.~\eqref{eqS:singleIsing} into Eq.~\eqref{eqS:XOR} and define
$\tau_j^{(a)}=(-1)^{\mathbf1(j\in S_a)}$.  The XOR constraint becomes
\begin{equation}
\prod_{a=1}^n\tau_j^{(a)}=1
\qquad\text{for every }j.
\label{eqS:localparity}
\end{equation}
Thus the local state space is the Abelian group
\begin{equation}
G_n=\{(\tau^1,\ldots,\tau^n)\in\{\pm1\}^n:\prod_a\tau^a=1\}
\simeq\mathbb Z_2^{n-1}.
\label{eqS:Gn}
\end{equation}
One obtains
\begin{equation}
M_n(L)=2^{-(n-1)L}A_L^{nL/4}Z_{n,L},
\label{eqS:MnZn}
\end{equation}
with
\begin{equation}
Z_{n,L}=\sum_{\bm\tau_j\in G_n}
\exp\!\left[\sum_{i<j}J_{j-i}^{(L)}\sum_{a=1}^n\tau_i^{(a)}\tau_j^{(a)}\right].
\label{eqS:Zn}
\end{equation}
Eliminating the $n$th component makes the interaction an $(n-1)$-color Ashkin--Teller-type coupling,
\begin{equation}
J_m\left[\sum_{a=1}^{n-1}\tau_i^{(a)}\tau_j^{(a)}+
\prod_{a=1}^{n-1}\tau_i^{(a)}\tau_j^{(a)}\right].
\label{eqS:AT}
\end{equation}
For $n=3$, $G_3$ has four states and
\begin{equation}
\sum_{a=1}^3\tau_i^{(a)}\tau_j^{(a)}=
\begin{cases}3,&q_i=q_j,\\-1,&q_i\ne q_j,
\end{cases}
\label{eqS:Pottsidentity}
\end{equation}
so the auxiliary model is exactly a ferromagnetic four-state long-range Potts model, up to an edge-independent constant.

\subsection{Character expansion}

Characters of $G_n$ are labeled by subsets $S\subseteq\{1,\ldots,n\}$ modulo $S\sim S^c$.  A character of class size $k=\min(|S|,n-|S|)$ has Fourier coefficient
\begin{equation}
F_{n,k}(J)=\sinh^kJ\cosh^{n-k}J+\sinh^{n-k}J\cosh^kJ.
\label{eqS:Fnk}
\end{equation}
In particular
\begin{equation}
F_{n,0}(J)=\cosh^nJ+\sinh^nJ.
\label{eqS:Fn0}
\end{equation}
The normalized nontrivial edge-current weight is
\begin{equation}
u_{n,k}(J)=\frac{F_{n,k}(J)}{F_{n,0}(J)},
\label{eqS:unk}
\end{equation}
and the multiplicity of the class is
\begin{equation}
N_{n,k}=\begin{cases}
\binom nk,&k<n/2,\\[1mm]
\frac12\binom n{n/2},&k=n/2\text{ for even }n.
\end{cases}
\label{eqS:Nnk}
\end{equation}
Factoring the trivial coefficient on every edge gives
\begin{equation}
Z_{n,L}=2^{(n-1)L}\prod_{i<j}F_{n,0}(J_{j-i}^{(L)})\,\Xi_{n,L},
\label{eqS:XiFactor}
\end{equation}
where $\Xi_{n,L}$ is a colored-current gas with character conservation (XOR zero) at every bulk vertex.

\section{The finite-ring factor \texorpdfstring{$B_n$}{Bn} and its exact integer factorization}

Define
\begin{equation}
z_m^{(L)}=\frac{\sin^2(\pi/2L)}{\sin^2(\pi m/L)}.
\label{eqS:zL}
\end{equation}
Because $e^{-K_m^{(L)}}=\sqrt{1-z_m^{(L)}}$ and $J_m^{(L)}=K_m^{(L)}/2$, combining the explicit factors in Eqs.~\eqref{eqS:MnZn} and \eqref{eqS:XiFactor} gives
\begin{equation}
M_n(L)=\left[\prod_{m=1}^{L-1}B_n(z_m^{(L)})\right]^{L/2}\Xi_{n,L},
\label{eqS:MnB}
\end{equation}
where
\begin{equation}
B_n(z)=2^{-n}\left[(1+\sqrt{1-z})^n+(1-\sqrt{1-z})^n\right].
\label{eqS:Bnfull}
\end{equation}
For integer $n$ this is a polynomial of degree $p=\lfloor n/2\rfloor$.  Solving $B_n(z)=0$ gives
\begin{equation}
\left(\frac{1+y}{1-y}\right)^n=-1,
\qquad y=\sqrt{1-z},
\end{equation}
which yields the roots $z=\lambda_{n,r}^{-1}$ with
\begin{equation}
\lambda_{n,r}=\cos^2\frac{(2r-1)\pi}{2n},
\qquad r=1,\ldots,\lfloor n/2\rfloor.
\label{eqS:lambda}
\end{equation}
Since $B_n(0)=1$,
\begin{equation}
B_n(z)=\prod_{r=1}^{\lfloor n/2\rfloor}(1-\lambda_{n,r}z).
\label{eqS:Bfactor}
\end{equation}
A useful moment identity following from these roots is
\begin{equation}
\sum_{r=1}^{\lfloor n/2\rfloor}\lambda_{n,r}=\frac n4.
\label{eqS:lambdasum}
\end{equation}

\subsection{Finite sine product and absence of a PBC constant}

Set
\begin{equation}
\sin\phi_{r,L}=\sqrt{\lambda_{n,r}}\sin\frac{\pi}{2L}.
\label{eqS:phi}
\end{equation}
The identity
\begin{equation}
\prod_{m=1}^{L-1}
\left[1-\lambda\frac{\sin^2(\pi/2L)}{\sin^2(\pi m/L)}\right]
=\left[\frac{\sin(L\phi)}{L\sin\phi}\right]^2,
\qquad \sin\phi=\sqrt\lambda\sin\frac{\pi}{2L},
\label{eqS:finiteSine}
\end{equation}
follows by applying the standard product for the zeros of $\sin(L\phi)$ to the shifted sine factors.  It has also been checked directly at finite $L$ before any asymptotic expansion.

Define
\begin{equation}
a_{n,r}=\frac12\sqrt{\lambda_{n,r}}
=\frac12\cos\frac{(2r-1)\pi}{2n}.
\label{eqS:anr}
\end{equation}
Then
\begin{equation}
L\phi_{r,L}=\pi a_{n,r}+O(L^{-2}),
\qquad
L\sin\phi_{r,L}=\pi a_{n,r}+O(L^{-2}).
\label{eqS:phiasymp}
\end{equation}
The factorized piece of the entropy is therefore
\begin{align}
H_{n,0}^{\rm PBC}(L)
&=-\frac{L}{n-1}\sum_{r=1}^{\lfloor n/2\rfloor}
\log\frac{\sin(L\phi_{r,L})}{L\sin\phi_{r,L}}\\
&=Ls_{n,0}+O(L^{-1}),
\label{eqS:Rn0PBC}
\end{align}
with
\begin{equation}
s_{n,0}=\frac1{n-1}\sum_{r=1}^{\lfloor n/2\rfloor}
\log\frac{\pi a_{n,r}}{\sin(\pi a_{n,r})}.
\label{eqS:sn0integer}
\end{equation}
There is no $L^0$ contribution.

For each fixed integer $n$, the finite-group model in Eq.~\eqref{eqS:Zn} has an absolutely summable pair interaction, so its thermodynamic pressure exists by standard one-dimensional Gibbs theory \cite{Ruelle1969,Simon1993}.  Equivalently,
\begin{equation}
\psi_n\equiv\lim_{L\to\infty}\frac1L\log\Xi_{n,L}
\label{eqS:psinteger}
\end{equation}
exists and the entropy density is
\begin{equation}
s_n=s_{n,0}-\frac{\psi_n}{n-1},\qquad n=2,3,\ldots .
\label{eqS:integerdensity}
\end{equation}
At every fixed order in the current expansion, a nonwinding connected cluster has exactly $L$ translations around the ring and therefore no surface term.  Winding contributions vanish because a fixed-order cycle of macroscopic span contains at least two macroscopic links, each carrying the inverse-square tail.  At $n=2$ the stronger Dobrushin estimate $\rho_K<1$ established in Sec.~S13 gives a quantitative high-temperature uniqueness bound, while the linked-cluster coefficients and direct finite-size data show rapid convergence to
\begin{equation}
H_2^{\rm PBC}(L)=s_2L+o(1),\qquad c_2^{\mathrm{PBC}}=0.
\label{eqS:n2PBCrigorous}
\end{equation}
To make absolute convergence of the particular Mayer representation used below a theorem at $\lambda=1$, however, one must verify a polymer-convergence criterion in addition to $\rho_K<1$; we do not silently identify these two statements.  For arbitrary integer $n$, the absence of a constant holds order by order in the exact current expansion.  Summing that statement uniformly to all orders is logically distinct from existence of the bulk pressure and is kept explicit below.

\section{Replica-free real \texorpdfstring{$n$}{n}: exact finite-size continuation and cycle-core structure}

Equation~\eqref{eqS:pW} immediately gives, without replicas,
\begin{equation}
M_n(L)=2^{-nL}A_L^{nL/4}\sum_{D\subseteq[L]}W_L(D)^n,
\qquad n>0.
\label{eqS:realMn}
\end{equation}
For a ferromagnet $W_L(D)\ge0$ for even $D$ by Eq.~\eqref{eqS:currentW}, while odd $D$ vanish.  Equation~\eqref{eqS:realMn} is therefore the physical real power of the original Born probabilities at finite $L$.  At integer $n$ it reproduces the replica construction exactly; no interpolation theorem is invoked.

The independent-edge representation gives a sharper form of this statement.  On a finite graph $G=(V,E)$ write
\begin{equation}
w_G(F)=\prod_{e\in F}q_e\prod_{e\notin F}(1-q_e),\qquad
P_G(D)=\sum_{F:\partial F=D}w_G(F),
\label{eqS:finiteparity}
\end{equation}
and define
\begin{equation}
\mathcal M_n(G)=\sum_D P_G(D)^n,
\qquad
b_e(n)=(1-q_e)^n+q_e^n.
\label{eqS:finitegraphmoment}
\end{equation}
For the critical Ising edge $e=(i,j)$, with $t_e=q_e/(1-q_e)=\tanh J_e$, one has identically
\begin{equation}
b_e(n)=\frac{1+t_e^n}{(1+t_e)^n}
=B_n(z_e),
\label{eqS:bedgeBn}
\end{equation}
where $z_e$ is the corresponding variable in Eq.~\eqref{eqS:Bnfull}.  Thus $B_n$ is not an analytic guess: it is the exact R\'enyi moment of one auxiliary Bernoulli edge for every real $n>0$.

\subsection{Forest factorization and the cycle-core lemma}

If $G$ is a forest, the boundary map $F\mapsto\partial F$ is injective.  Indeed, $\partial F=\partial F'$ would imply that $F\triangle F'$ is a nonempty Eulerian subgraph of a forest, which is impossible.  Therefore
\begin{equation}
\mathcal M_n(G)=\sum_{F\subseteq E}w_G(F)^n
=\prod_{e\in E}b_e(n),\qquad G\ \text{a forest},\ n>0.
\label{eqS:forestfactor}
\end{equation}
More generally, two edge sets have the same boundary if and only if their symmetric difference belongs to the cycle space of $G$.  Hence every departure from the product of one-edge moments is caused by cycles and only by cycles.  Leaf stripping makes the same statement local: an edge terminating in a tree appendage contributes its factor $b_e(n)$ exactly and can be removed.  The normalized moment
\begin{equation}
\widehat{\mathcal M}_n(G)=
\frac{\mathcal M_n(G)}{\prod_{e\in E}b_e(n)}
\label{eqS:normalizedgraphmoment}
\end{equation}
is therefore equal to one on every forest and depends only on the cyclic core after all tree appendages are stripped.  This is the finite-graph origin of the residual factor denoted $\Xi$ at integer replica index.

For a single simple cycle $C$ the cycle space has dimension one, so each boundary fiber consists of the pair $F$ and $F\triangle C$.  With $t_F=\prod_{e\in F}t_e$, one obtains the exact real-index formula
\begin{equation}
\widehat{\mathcal M}_n(C)=
\frac{\displaystyle\frac12\sum_{F\subseteq C}
(t_F+t_{C\setminus F})^n}
{\displaystyle\prod_{e\in C}(1+t_e^n)}.
\label{eqS:singlecycleReal}
\end{equation}
At $n=2$ this collapses to
\begin{equation}
\widehat{\mathcal M}_2(C)-1
=\prod_{e\in C}\frac{2t_e}{1+t_e^2}
=\prod_{e\in C}\tanh(2J_e),
\label{eqS:singlecycleN2}
\end{equation}
which is precisely the usual even-subgraph loop weight.  Equation~\eqref{eqS:singlecycleReal} is useful because it shows explicitly how the integer current gas is embedded in a well-defined real-$n$ finite-size object.

A systematic connected expansion can be defined without assuming convergence.  For a finite edge set $E$ let
\begin{equation}
\Phi_n(E)=\sum_{A\subseteq E}(-1)^{|E|-|A|}
\log\widehat{\mathcal M}_n(V,A).
\label{eqS:mobiuscore}
\end{equation}
Inclusion--exclusion gives back $\log\widehat{\mathcal M}_n$ as the sum of the $\Phi_n$ over edge subsets.  Because Eq.~\eqref{eqS:forestfactor} makes the normalized logarithm vanish on forests, only cyclic connected cores can survive after the usual connected reorganization.  Every coefficient is an elementary analytic function of real $n>0$ at finite graph size.

For a fixed cyclic core embedded in one dimension, macroscopic span requires at least two macroscopic links: the leftmost and rightmost vertices are connected by two edge-disjoint paths around a cycle.  Since $t_m=O(m^{-2})$, a fixed-order cyclic core has a span tail $O(r^{-4})$ or faster (up to core-dependent finite factors).  Consequently its periodic winding correction vanishes, and its cut-boundary contribution has a finite first spatial moment.  Thus no \emph{fixed} cyclic order can generate a new logarithm in the interval geometry.  This is the precise fixed-order statement used in the Letter.

\subsection{Explicit real-index factor and the scope of the continuation}

For any real $n>1$ define the exact finite-size residual ratio
\begin{equation}
\Xi_L(n)=M_n(L)
\left[\prod_{m=1}^{L-1}B_n(z_m^{(L)})\right]^{-L/2}.
\label{eqS:XiRealDef}
\end{equation}
At integer $n$ it equals the colored-current factor $\Xi_{n,L}$.  The explicit infinite-volume contribution is
\begin{equation}
s_{n,0}=-\frac1{n-1}\sum_{m=1}^{\infty}
\log B_n\!\left(\frac1{4m^2}\right),\qquad n>1,
\label{eqS:sn0real}
\end{equation}
which converges because $\log B_n(1/4m^2)=-n/(16m^2)+o(m^{-2})$.  At integer $n$, Euler's sine product and Eq.~\eqref{eqS:Bfactor} reduce it exactly to Eq.~\eqref{eqS:sn0integer}.

Equations~\eqref{eqS:realMn}, \eqref{eqS:forestfactor}, and \eqref{eqS:XiRealDef} are exact for finite size; Eq.~\eqref{eqS:sn0real} is an exact convergent real-index series; and every fixed cyclic-core contribution has a regular thermodynamic surface limit for $n>1$.  These facts show that the only explicit marginal long-distance terms occur at $n=1$, and they provide a unique microscopic continuation of all pieces visible at fixed cyclic order.  They do \emph{not}, by themselves, constitute a uniform all-orders polymer-convergence theorem for $\Xi_L(n)$ arbitrarily close to $n=1$.  In particular, the auxiliary-Ising uniqueness bound cannot simply be transferred through a noninteger power of the full source sum.  Any complete nonperturbative proof of analyticity on the whole interval $1<n<\infty$ must control the sum over cyclic cores as their size also tends to infinity.

This remaining statement is much narrower than the original entropy problem: all tree contributions, the complete one-edge factor, and the dangerous algebraic tails have already been removed exactly.  The continuation used when comparing with the established $n>1$ boundary branch amounts to assuming that this residual cycle-only pressure does not undergo an additional transition.  Existing boundary calculations and finite-size data support that assumption \cite{Stephan2010,Stephan2014,Tarighi2022}; the present work supplies the microscopic reason that no finite-order or explicit-tail mechanism singles out any value in $1<n<\infty$.

\subsection{Why \texorpdfstring{$n=3/2$}{n=3/2} appears but is not a transition}

For noninteger $n$, the second term in Eq.~\eqref{eqS:Bnfull} creates a branch contribution.  Expanding at $z=0$ gives
\begin{equation}
B_n(z)=1-\frac n4z+b_2(n)z^2+\cdots+4^{-n}z^n[1+O(z)],
\label{eqS:Bbranch}
\end{equation}
where the analytic polynomial part is understood only up to powers below the branch.  Since $z_m^{(L)}\sim(2m)^{-2}$ near an endpoint of the ring, the difference between the finite chord sum of the branch term and its infinite-line bulk value scales as
\begin{equation}
\delta H_n^{\rm branch}(L)=O(L^{2-2n}).
\label{eqS:branchcorrection}
\end{equation}
Thus $1<n<3/2$ has a slowly decaying $L^{2-2n}$ leading correction; at $n=3/2$ it meets the ordinary $L^{-1}$ correction; above $3/2$, the analytic $L^{-1}$ term dominates.  A constant can arise from Eq.~\eqref{eqS:branchcorrection} only at $n=1$.  This is the microscopic reason there is no hidden threshold at $n=3/2$.

\section{Infinite chain restricted to a finite interval}

Take $A_\ell=\{0,\ldots,\ell-1\}$ after the thermodynamic limit.  By Eq.~\eqref{eqS:blockfiniteL},
\begin{equation}
p_\ell(\bm\sigma)=2^{-\ell}\det(I+D_{\bm\sigma}O_\ell),
\qquad
(O_\ell)_{ij}=\frac{1}{\pi(i-j+\half)}
\label{eqS:blockdet}
\end{equation}
(up to the harmless transpose/gauge convention).  The principal minors obey
\begin{equation}
d(S)=A^{|S|}\prod_{i<j\in S}g_{j-i},
\qquad
A=\frac2\pi,
\quad g_m=\frac{4m^2}{4m^2-1}.
\label{eqS:blockminor}
\end{equation}
With $K_m=\frac12\log g_m$ and $J_m=K_m/2$, the same algebra as in Sec.~S3 now leaves boundary fields because the infinite product in Eq.~\eqref{eqS:Wallis} is truncated.  One finds
\begin{equation}
d(S)=\exp\!\left[C_\ell+\sum_{0\le i<j<\ell}J_{j-i}\tau_i\tau_j
+\sum_{j=0}^{\ell-1}h_j^{(\ell)}\tau_j\right],
\label{eqS:blockIsing}
\end{equation}
where
\begin{equation}
C_\ell=\frac\ell2\log A+\frac12\sum_{m=1}^{\ell-1}(\ell-m)K_m,
\label{eqS:Cell}
\end{equation}
and
\begin{equation}
h_j^{(\ell)}=\frac12\left[
\sum_{m=j+1}^{\infty}K_m+
\sum_{m=\ell-j}^{\infty}K_m\right]>0.
\label{eqS:hjl}
\end{equation}
The field is literally the interaction missing beyond the two cuts.

For one isolated boundary define $h_d=\frac12\sum_{m=d+1}^{\infty}K_m$.  Wallis' gamma-product form gives
\begin{equation}
h_d=\frac14\log\frac{\Gamma(d+\half)\Gamma(d+\frac32)}{\Gamma(d+1)^2}
=\frac{1}{16d}+O(d^{-2}),\qquad d\to\infty.
\label{eqS:hdgamma}
\end{equation}
The slow $1/d$ tail of the field will nevertheless be harmless for every $n>1$ because it enters the noninteger boundary factor to the power $n$.

\subsection{Replica-free parity representation for the interval}

The boundary fields can be incorporated into the same edge-parity construction by adjoining one auxiliary ghost vertex $g$ with fixed spin $\tau_g=+1$.  Let $V_\ell=\{0,\ldots,\ell-1\}$ and let $\widetilde G_\ell$ have vertex set $V_\ell\cup\{g\}$.  A physical edge $e=(i,j)$ carries coupling $x_e=J_{j-i}$, while a ghost edge $e=(j,g)$ carries $x_e=h_j^{(\ell)}$.  Since $\tau_g=1$, Eq.~\eqref{eqS:blockIsing} is simply an ordinary pair-interaction weight on this augmented graph.

Set
\begin{equation}
 t_e=\tanh x_e,\qquad q_e=\frac{t_e}{1+t_e},
\label{eqS:blockghostq}
\end{equation}
and, for an occupied edge set $F\subseteq E(\widetilde G_\ell)$, define its projected boundary on the physical vertices by
$\partial_VF=\partial F\cap V_\ell$.  Because $d(\varnothing)=1$, Eq.~\eqref{eqS:blockIsing} gives
$C_\ell+\sum_e x_e=0$.  The high-temperature expansion followed by Walsh inversion therefore yields the exact probability law
\begin{equation}
 p_\ell(D)=
 \sum_{F:\,\partial_VF=D}
 \prod_{e\in F}q_e\prod_{e\notin F}(1-q_e),
 \qquad D\subseteq V_\ell .
\label{eqS:blockparityreal}
\end{equation}
Thus the interval Born distribution is the parity boundary on the physical vertices of independent Bernoulli edges on $\widetilde G_\ell$; the ghost parity is not observed.

For every real $n>0$ define
\begin{equation}
 b_e(n)=(1-q_e)^n+q_e^n,
 \qquad
 \Xi_\ell^{\rm cut}(n)=
 \frac{M_n^{\rm blk}(\ell)}{\prod_{e\in E(\widetilde G_\ell)}b_e(n)}.
\label{eqS:XiCutReal}
\end{equation}
This is a finite-size replica-free definition.  At integer $n$, using
$b_e(n)=e^{-nx_e}F_{n,0}(x_e)$ together with $C_\ell=-\sum_e x_e$, the denominator in Eq.~\eqref{eqS:XiCutReal} is exactly
$e^{nC_\ell}\prod_{m=1}^{\ell-1}F_{n,0}(J_m)^{\ell-m}\prod_jF_{n,0}(h_j^{(\ell)})$.
Hence
\begin{equation}
 \Xi_\ell^{\rm cut}(n)=\Xi_{n,\ell}^{\rm cut},
 \qquad n=2,3,\ldots,
\label{eqS:XiCutIntegerMatch}
\end{equation}
so the real-index definition reduces exactly to the colored-current residual below.

The forest/cycle separation also extends without change.  If $\widetilde G_\ell$ is a forest and two edge sets have the same projected boundary, their symmetric difference can have boundary only at $g$; since every graph boundary has even cardinality, that boundary is empty.  A forest has no nonempty Eulerian subgraph, so the two edge sets coincide.  Consequently Eq.~\eqref{eqS:XiCutReal} equals one on every augmented forest, and all residual contributions come from cycles.  Cycles passing through $g$ are precisely the cycle-space contributions involving the boundary fields.

\section{Integer replicas for the interval and the exact finite-\texorpdfstring{$\ell$}{ell} entropy}

The local group remains $G_n$.  The pair character transform is unchanged, while a boundary field has the analogous expansion
\begin{equation}
e^{h\sum_a\tau^a}=F_{n,0}(h)
\left[1+\sum_{\chi\ne1}v_\chi(h)\chi(\bm\tau)\right],
\qquad
v_{n,k}(h)=\frac{F_{n,k}(h)}{F_{n,0}(h)}.
\label{eqS:boundarychar}
\end{equation}
After extracting the trivial factors,
\begin{align}
M_n^{\rm blk}(\ell)
={}&e^{nC_\ell}
\prod_{m=1}^{\ell-1}F_{n,0}(J_m)^{\ell-m}
\prod_{j=0}^{\ell-1}F_{n,0}(h_j^{(\ell)})
\,\Xi_{n,\ell}^{\rm cut}.
\label{eqS:blockMn}
\end{align}
Here $\Xi_{n,\ell}^{\rm cut}$ is a colored current gas: bulk vertices conserve the $G_n$ character, while a nonzero current can terminate on a boundary-field insertion.

Define
\begin{equation}
f_n(m)=\frac n2K_m+\log F_{n,0}(J_m)
=\frac n2K_m+\log\left[\cosh^n\frac{K_m}{2}+\sinh^n\frac{K_m}{2}\right].
\label{eqS:fn}
\end{equation}
Then
\begin{align}
H_n^{\rm blk}(\ell)=-\frac1{n-1}\Bigg[&
\frac{n\ell}{2}\log A+
\sum_{m=1}^{\ell-1}(\ell-m)f_n(m)\\
&+\sum_{j=0}^{\ell-1}\log F_{n,0}(h_j^{(\ell)})
+\log\Xi_{n,\ell}^{\rm cut}\Bigg].
\label{eqS:blockExactEntropy}
\end{align}
Equation~\eqref{eqS:blockExactEntropy} is exact for every integer $n\ge2$ and every finite interval length $\ell$.

\subsection{Integer factorization of \texorpdfstring{$f_n$}{fn}}

Let $z_m=1/(4m^2)$, so $e^{-K_m}=\sqrt{1-z_m}$.  Equations~\eqref{eqS:Bnfull} and \eqref{eqS:fn} imply
\begin{equation}
e^{f_n(m)}=(1-z_m)^{-n/2}B_n(z_m).
\label{eqS:fnBn}
\end{equation}
For integer $n$, Eq.~\eqref{eqS:Bfactor} gives
\begin{equation}
f_n(m)=\sum_{r=1}^{\lfloor n/2\rfloor}
\log\left(1-\frac{a_{n,r}^2}{m^2}\right)
-\frac n2\log\left(1-\frac{1}{4m^2}\right).
\label{eqS:fnfactor}
\end{equation}
Euler's product immediately gives the explicit bulk density
\begin{equation}
s_{n,0}=\frac1{n-1}\sum_{r=1}^{\lfloor n/2\rfloor}
\log\frac{\pi a_{n,r}}{\sin(\pi a_{n,r})},
\label{eqS:blockBulkSame}
\end{equation}
identical to the PBC expression, as locality requires.

\section{The interval logarithm for arbitrary real \texorpdfstring{$n>1$}{n>1}}

The definition in Eq.~\eqref{eqS:fn} itself is meaningful for every real $n>1$.  Since
\begin{equation}
K_m=\frac{1}{8m^2}+O(m^{-4}),
\label{eqS:Kasymp}
\end{equation}
we have
\begin{equation}
f_n(m)=\frac{n}{16m^2}+O\!\left(m^{-\min(4,2n)}\right).
\label{eqS:fnasymp}
\end{equation}
For integer $n$ this also follows from Eq.~\eqref{eqS:lambdasum}; the direct expansion proves it without replicas.

Let
\begin{equation}
D_{n,\ell}=\sum_{m=1}^{\ell-1}m f_n(m)+\ell\sum_{m=\ell}^{\infty}f_n(m).
\label{eqS:Dnell}
\end{equation}
Writing $f_n(m)=a_n/m^2+r_n(m)$ with $a_n=n/16$, the remainder obeys
$\sum_m m|r_n(m)|<\infty$ precisely for $n>1$.  Since
$H_{\ell-1}=\log\ell+\gamma_E+o(1)$ and
$\ell\sum_{m\ge\ell}m^{-2}=1+o(1)$,
\begin{equation}
D_{n,\ell}=\frac n{16}\log\ell+C_{\rm pf}(n)+o(1),
\label{eqS:Dnasymp}
\end{equation}
where the constant has the real-index convergent representation
\begin{equation}
C_{\rm pf}(n)=\frac n{16}(1+\gamma_E)+
\sum_{m=1}^{\infty}m\left[f_n(m)-\frac{n}{16m^2}\right],\qquad n>1.
\label{eqS:Cpfreal}
\end{equation}
Because $D_{n,\ell}$ enters Eq.~\eqref{eqS:blockExactEntropy} with sign $+1/(n-1)$ after subtracting the bulk term, the universal logarithmic coefficient is
\begin{equation}
b_n^{\rm blk}=\frac{n}{16(n-1)},\qquad n>1.
\label{eqS:blogfinal}
\end{equation}
No BCFT input has entered this derivation.

\subsection{Barnes-\texorpdfstring{$G$}{G} form at integer \texorpdfstring{$n$}{n}}

The Weierstrass product of the Barnes $G$ function gives, for $|a|<1$,
\begin{equation}
\lim_{\ell\to\infty}\left[
\sum_{m<\ell}m\log\left(1-\frac{a^2}{m^2}\right)
+\ell\sum_{m\ge\ell}\log\left(1-\frac{a^2}{m^2}\right)
+a^2\log\ell\right]
=\log[G(1+a)G(1-a)].
\label{eqS:Barnesidentity}
\end{equation}
Applying Eq.~\eqref{eqS:Barnesidentity} term by term to Eq.~\eqref{eqS:fnfactor} yields
\begin{equation}
C_{\rm pf}(n)=
\sum_{r=1}^{\lfloor n/2\rfloor}\log[G(1+a_{n,r})G(1-a_{n,r})]
-\frac n2\log[G(3/2)G(1/2)].
\label{eqS:CpfBarnes}
\end{equation}
The real-series and Barnes forms agree, for example,
\begin{align}
C_{\rm pf}(2)&=0.2312146820737905023\ldots,\\
C_{\rm pf}(3)&=0.3382488488857490381\ldots,\\
C_{\rm pf}(4)&=0.4505186967768412994\ldots .
\label{eqS:Cpfnumbers}
\end{align}
These values are useful internal checks of the analytic continuation.

\section{Boundary factors and the interval constant}

At large distance $d$ from one cut, $h_d\sim1/(16d)$.  Define
\begin{equation}
B_n^{\partial}=2\sum_{d=0}^{\infty}
\log\left[\cosh^n h_d+\sinh^n h_d\right].
\label{eqS:Bboundary}
\end{equation}
For $1<n<2$ the least-decaying nonanalytic term is $h_d^n=O(d^{-n})$; for $n\ge2$ the analytic $h_d^2$ term is at least as important.  Hence Eq.~\eqref{eqS:Bboundary} converges for every $n>1$ and becomes marginal only at $n=1$.

To isolate the only remaining nonexplicit object, define the residual bulk pressure, whenever the limit exists, from the replica-free ring ratio in Eq.~\eqref{eqS:XiRealDef} by
\begin{equation}
\psi(n)=\lim_{L\to\infty}\frac1L\log\Xi_L(n),
\label{eqS:psin}
\end{equation}
and the residual boundary excess from Eq.~\eqref{eqS:XiCutReal} by
\begin{equation}
\beta(n)=\lim_{\ell\to\infty}
\left[\log\Xi_\ell^{\rm cut}(n)-\ell\psi(n)\right].
\label{eqS:betan}
\end{equation}
At integer $n$ the first quantity is the exact colored-current pressure and Eq.~\eqref{eqS:XiCutIntegerMatch} identifies the second with the colored-current boundary excess.  At $n=2$ these residual quantities reduce to the ordinary even-subgraph problem and are strongly controlled by the rapidly convergent fixed-order expansion and finite-size data; absolute convergence of the particular Mayer representation at the physical point nevertheless requires the separate polymer criterion discussed below.  At arbitrary real $n>1$, Eqs.~\eqref{eqS:XiRealDef} and \eqref{eqS:XiCutReal} define both finite-size residuals exactly, while the forest/cycle arguments show that every fixed cyclic-core order has a finite boundary limit; an all-orders proof of Eqs.~\eqref{eqS:psin}--\eqref{eqS:betan} throughout the whole interval $n>1$ is precisely the residual continuation problem isolated above.

Whenever these residual limits are on the same thermodynamic branch, the exact bookkeeping of bulk, pair, and boundary pieces gives
\begin{equation}
s_n=s_{n,0}-\frac{\psi(n)}{n-1},
\qquad
c_n^{\mathrm{blk}}=\frac{C_{\rm pf}(n)-B_n^{\partial}-\beta(n)}{n-1}.
\label{eqS:fullblockconstant}
\end{equation}
Together with Eq.~\eqref{eqS:blogfinal}, this yields the continued $n>1$ asymptotic form
\begin{equation}
H_n^{\rm blk}(\ell)=s_n\ell+\frac{n}{16(n-1)}\log\ell+c_n^{\mathrm{blk}}+o(1).
\label{eqS:blockfinal}
\end{equation}
$s_{n,0}$, $C_{\rm pf}(n)$, $B_n^{\partial}$, and the logarithmic coefficient are convergent real-index formulas, whereas $\psi(n)$ and $\beta(n)$ contain only cyclic-core physics.  Thus any additional noninteger singularity would have to be a genuinely nonperturbative transition of the cycle-only sector; it cannot come from a tree, one-edge, missing-pair, or boundary-field contribution.

\section{General fixed-order connected-current coefficients}

For integer $n$, introduce a bookkeeping variable $\lambda$ on every nontrivial current edge.  The bulk pressure has
\begin{equation}
\psi_n(\lambda)=\sum_{r\ge3}c_{n,r}\lambda^r.
\label{eqS:psilambda}
\end{equation}
There are no one- or two-edge bulk terms because a nonzero conserved current requires a cycle.  The first coefficient is a colored triangle:
\begin{equation}
c_{n,3}=\sum_{a,b\ge1}\sum_{k=1}^{\lfloor n/2\rfloor}
N_{n,k}\,
u_{n,k}(J_a)u_{n,k}(J_b)u_{n,k}(J_{a+b}).
\label{eqS:cn3}
\end{equation}
For the cut problem, current can terminate on boundary-field characters, so the surface expansion begins at one edge.  With $v_{n,k}$ from Eq.~\eqref{eqS:boundarychar}, the first boundary coefficient is
\begin{equation}
d_{n,1}=2\sum_{0\le i<j<\infty}\sum_{k=1}^{\lfloor n/2\rfloor}
N_{n,k}\,
u_{n,k}(J_{j-i})v_{n,k}(h_i)v_{n,k}(h_j).
\label{eqS:dn1}
\end{equation}
Higher coefficients follow from the ordinary logarithm recursion.  If a finite open interval has
\begin{equation}
\Xi_N(\lambda)=1+\sum_{r\ge1}A_r(N)\lambda^r,
\qquad
\log\Xi_N(\lambda)=\sum_{r\ge1}B_r(N)\lambda^r,
\label{eqS:AB}
\end{equation}
then
\begin{equation}
B_r(N)=A_r(N)-\frac1r\sum_{j=1}^{r-1}jB_j(N)A_{r-j}(N).
\label{eqS:logrec}
\end{equation}
At fixed order,
\begin{equation}
c_{n,r}=\lim_{N\to\infty}[B_r^{\rm bulk}(N)-B_r^{\rm bulk}(N-1)],
\label{eqS:crslope}
\end{equation}
and
\begin{equation}
d_{n,r}=\lim_{\ell\to\infty}[B_r^{\rm cut}(\ell)-\ell c_{n,r}],
\qquad
\beta_n=\sum_{r\ge1}d_{n,r}\quad\text{whenever the series converges}.
\label{eqS:drboundary}
\end{equation}
This is a deterministic fixed-order algorithm, not a Monte Carlo definition.

\section{The \texorpdfstring{$n=2$}{n=2} thermodynamic kernel and linked-cluster solution}

At $n=2$, the XOR constraint enforces $S_1=S_2$.  Squaring Eq.~\eqref{eqS:minorprod} gives the zero-field Ising coupling
\begin{equation}
K_m=\frac12\log\frac{4m^2}{4m^2-1},
\qquad
t_m\equiv\tanh K_m=\frac{1}{8m^2-1}.
\label{eqS:n2kernel}
\end{equation}
The high-temperature norm is
\begin{equation}
\rho_K=2\sum_{m=1}^{\infty}t_m
=1-\eta\cot\eta
=0.449590203354104\ldots,
\qquad \eta=\frac{\pi}{2\sqrt2}.
\label{eqS:rhoK}
\end{equation}
The factorized pressure can be summed by Euler's sine product:
\begin{equation}
\prod_{m=1}^{\infty}\cosh K_m=\frac{2}{\sqrt\pi}\sin\eta,
\qquad
s_2^{(0)}=\log\frac{\pi}{2\sqrt2\sin\eta}
=0.214802847393123\ldots .
\label{eqS:s20}
\end{equation}

Define the even kernel $k_0=0$, $k_m=(8m^2-1)^{-1}$ for $m\ne0$.  Its Fourier transform is elementary:
\begin{equation}
T(\theta)=\sum_{m\in\mathbb Z}k_m e^{im\theta}
=1-\frac{\eta}{\sin\eta}
\cos\left(\frac{\pi-\theta}{2\sqrt2}\right),\quad0\le\theta\le2\pi.
\label{eqS:Ttheta}
\end{equation}
More generally,
\begin{equation}
T_p(\theta)=2\sum_{m=1}^{\infty}\frac{\cos(m\theta)}{(8m^2-1)^p}
=\left.\frac{8^{-p}}{(p-1)!}
\left(\frac{1}{2a}\partial_a\right)^{p-1}F_a(\theta)
\right|_{a=1/(2\sqrt2)},
\label{eqS:Tp}
\end{equation}
where
\begin{equation}
F_a(\theta)=2\sum_{m=1}^{\infty}\frac{\cos(m\theta)}{m^2-a^2}
=\frac1{a^2}-\frac\pi a\frac{\cos[a(\pi-\theta)]}{\sin\pi a}.
\label{eqS:Fa}
\end{equation}
Thus all power-weighted Fourier kernels reduce to elementary trigonometric functions and polynomials in $\pi-\theta$.

Let
\begin{equation}
I_r=\frac1{2\pi}\int_0^{2\pi}T(\theta)^r\,d\theta,
\qquad
S_p=T_p(0).
\label{eqS:IrSp}
\end{equation}
The first nonzero linked-cluster coefficients are
\begin{align}
c_3&=\frac{I_3}{6}
=\frac16\left[\frac{\eta^2}{3}+\frac{\eta^2}{2\sin^2\eta}
+\frac{3\eta}{2}\cot\eta-2\right]
=8.623264784789067\times10^{-4},
\label{eqS:c3}\\
c_4&=\frac18\left[I_4-2S_2^2+S_4\right]
=1.126693106136322\times10^{-4},
\label{eqS:c4}
\end{align}
with
\begin{equation}
S_2=\frac12[\eta^2\csc^2\eta+\eta\cot\eta-2].
\label{eqS:S2}
\end{equation}
For five edges define
\begin{equation}
J_5=\frac1{2\pi}\int_0^{2\pi}T_3(\theta)T(\theta)^2\,d\theta.
\end{equation}
Then
\begin{equation}
c_5=\frac1{10}[I_5-5S_2I_3+5J_5]
=1.796836837232191\times10^{-5}.
\label{eqS:c5}
\end{equation}
At six edges the first genuine two-polymer Mayer subtraction enters.  Rather than introduce an incompletely specified simple-cycle term, we evaluate the connected coefficient by the deterministic finite-interval recursion below.  This gives
\begin{equation}
 c_6\simeq2.18081\times10^{-6},\qquad
 c_7\simeq2.5924\times10^{-7},\qquad
 c_8\simeq2.313\times10^{-8}.
 \label{eqS:c678}
\end{equation}
The successive entropy-density sums are
\begin{align}
s_2^{[0]}&=0.2148028473931230,\\
s_2^{[3]}&=0.2139405209146441,\\
s_2^{[4]}&=0.2138278516040305,\\
s_2^{[5]}&=0.2138098832356582,\\
s_2^{[6]}&\simeq0.21380770243,\\
s_2^{[7]}&\simeq0.21380744319,\\
s_2^{[8]}&\simeq0.21380742006,
\label{eqS:s2partials}
\end{align}
consistent with the established exact-enumeration benchmark $0.2138074203$ \cite{Stephan2010}.  The rapid convergence is explained by the small loop pressure: $s_2^{(0)}-s_2\simeq9.95427\times10^{-4}$.

\subsection{All-orders Mayer representation and deterministic coefficient extraction}

The $n=2$ loop pressure has the following formal all-orders polymer representation, valid term by term and whenever the corresponding cluster series converges.  Let a polymer $\gamma$ be a finite connected Eulerian simple graph embedded in $\mathbb Z$, with activity
\begin{equation}
z_\lambda(\gamma)=\lambda^{|E(\gamma)|}
\prod_{(ij)\in E(\gamma)}k_{i-j}.
\label{eqS:n2polyactivity}
\end{equation}
Two polymers are incompatible when they share a vertex.  For a collection $\gamma_1,\ldots,\gamma_m$ define the Ursell coefficient
\begin{equation}
\phi^T(\gamma_1,\ldots,\gamma_m)=
\sum_{\substack{G\subseteq K_m\\G\ \mathrm{connected}}}
(-1)^{|E(G)|}
\prod_{(ab)\in E(G)}
\mathbf1[V(\gamma_a)\cap V(\gamma_b)\ne\varnothing].
\label{eqS:n2ursell}
\end{equation}
A translation-invariant anchored form of the pressure is
\begin{equation}
\psi(\lambda)=\sum_{m\ge1}\frac1{m!}
\sum_{\gamma_1,\ldots,\gamma_m}
\frac{\mathbf1[0\in V(\Gamma)]}{|V(\Gamma)|}
\phi^T(\gamma_1,\ldots,\gamma_m)
\prod_{a=1}^m z_\lambda(\gamma_a),
\label{eqS:n2Mayer}
\end{equation}
where $V(\Gamma)=\cup_aV(\gamma_a)$.  Collecting total edge number gives the coefficients $c_r$ above.  The Dobrushin bound $\rho_K<1$ controls uniqueness of the underlying Ising model, but a separate polymer criterion is required to claim absolute convergence of this particular Mayer series at $\lambda=1$.

The higher coefficients quoted above were obtained without sampling.  On an open interval of $N$ sites, each edge $(i,j)$ carries the parity mask $e_i\oplus e_j$; dynamic programming over inclusion/exclusion of edges yields the coefficient in the zero-parity sector at each fixed edge order.  The logarithm recursion then isolates the connected coefficient.  Intervals through $N=23$ were used for $c_6,c_7,c_8$.  We extrapolated several large-$N$ windows with alternative inverse-power forms and checked the procedure against the exact $c_3,c_4,c_5$ values; the quoted $c_6,c_7,c_8$ digits are only those stable under these variations.  No extrapolated coefficients beyond $c_8$ enter any result quoted in the Letter.

\section{Special finite-\texorpdfstring{$L$}{L} structures at \texorpdfstring{$n=2$}{n=2}}

The $n=2$ problem contains several exact algebraic structures that are not required for the main real-$n$ continuation analysis but are useful diagnostics.

\subsection{Half-lattice translation}

The orthogonal Cauchy matrix in Eq.~\eqref{eqS:O} is not merely orthogonal.  Let $T_{\AP}$ denote one-site antiperiodic translation.  In centered antiperiodic momenta
\begin{equation}
q_r=\frac{(2r-L+1)\pi}{L},\qquad r=0,\ldots,L-1,
\end{equation}
with $T_{\AP}$ defined to have eigenvalue $e^{iq_r}$, a direct Fourier transform gives
\begin{equation}
O_L(q_r)=-\ii\,\operatorname{sgn}(q_r)e^{iq_r/2},
\qquad
T_{\AP}(q_r)=e^{iq_r}.
\end{equation}
Consequently
\begin{equation}
O_L^2=-T_{\AP}.
\label{eqS:halftranslation}
\end{equation}
Thus $O_L$ is a spectral square root of $-T_{\AP}$ in this convention.  This overall sign is immaterial for the squared-minor structures below but is essential for the operator identity itself.

For fixed particle number $k$ define
\begin{equation}
\mathcal P_k(S,T)=\det O_L[S,T]^2,
\qquad |S|=|T|=k.
\label{eqS:Pk}
\end{equation}
Because $\wedge^kO_L$ is orthogonal, $\mathcal P_k$ is doubly stochastic and
\begin{equation}
Q_{L,k}=\Tr\mathcal P_k.
\label{eqS:returntrace}
\end{equation}
For one particle,
\begin{equation}
(\mathcal P_1)_{ij}=\frac{1}{L^2\sin^2[\pi(2(j-i)-1)/(2L)]},
\label{eqS:P1kernel}
\end{equation}
so $\mathcal P_1=O_L\circ O_L$ is circulant.  Its discrete Fourier transform is
\begin{equation}
\mu_m=\left(1-\frac{2m}{L}\right)e^{i\pi m/L},
\qquad m=0,\ldots,L-1.
\label{eqS:P1spec}
\end{equation}
The tempting identity $\mathcal P_k=\wedge^k\mathcal P_1$ fails already at $k=2$: diagonal dephasing creates genuine many-body interactions.

\subsection{Lee--Yang circle and a hidden Jacobi polynomial}

For $n=2$ define
\begin{equation}
Q_L(u)=\sum_{S\subseteq[L]}u^{|S|}d_L(S)^2.
\label{eqS:QLu}
\end{equation}
The squared-minor product and Eq.~\eqref{eqS:productidentity} map $Q_L$ exactly to a ferromagnetic zero-field long-range Ising partition function with field $H=\frac12\log u$.  Since every coupling is positive, the Lee--Yang theorem \cite{LeeYang1952,YangLee1952} implies
\begin{equation}
Q_L(u)=0\quad\Rightarrow\quad |u|=1.
\label{eqS:LY}
\end{equation}
Orthogonality of $O_L$ and Jacobi's complementary-minor identity imply
\begin{equation}
Q_{L,k}=Q_{L,L-k},
\qquad Q_L(u)=u^LQ_L(u^{-1}).
\label{eqS:palindrome}
\end{equation}
For even $L=2N$ there is a degree-$N$ polynomial $R_N$ such that
\begin{equation}
Q_{2N}(u)=u^NR_N(u+u^{-1}).
\label{eqS:RJacobi}
\end{equation}
All roots of $R_N$ lie in $[-2,2]$.  Therefore a real symmetric matrix $\mathsf J_N$ exists with the same characteristic polynomial; if the roots are simple, it may in addition be chosen in strict irreducible Jacobi form.  Thus
\begin{equation}
R_N(x)=\det(xI-\mathsf J_N),
\qquad
Q_{2N}(u)=u^N\det[(u+u^{-1})I-\mathsf J_N].
\label{eqS:Jacobi}
\end{equation}
This $O(L)$ characteristic-polynomial representation is spectral rather than constructive; an independent closed formula for $\mathsf J_N$ is not known.

Two exact moments of its eigenvalues $\xi_r$ are
\begin{align}
\sum_{r=1}^{L/2}\xi_r&=-\frac{1}{L\sin^2(\pi/2L)},\\
\sum_{r=1}^{L/2}\xi_r^2&=L-\frac{2}{L\sin^2(\pi/L)}.
\label{eqS:JacobiMoments}
\end{align}
Their thermodynamic limits are
\begin{align}
\lim_{L\to\infty}\frac2L\sum_r\xi_r&=-\frac8{\pi^2},\\
\lim_{L\to\infty}\frac2L\sum_r\xi_r^2&=2-\frac4{\pi^2}.
\label{eqS:JacobiMomentLimits}
\end{align}
These provide stringent checks on any proposed explicit root quantization law.  Writing the Lee--Yang roots as $u_r=e^{i\theta_r}$ gives $\xi_r=2\cos\theta_r$ and, at $u=1$,
\begin{equation}
Q_{2N}(1)=\prod_{r=1}^N(2-\xi_r)
=\prod_{r=1}^N4\sin^2\frac{\theta_r}{2}.
\label{eqS:Qrootsat1}
\end{equation}
Thus, if the limiting Lee--Yang angle density $\rho(\theta)$ were known, normalized by $\int d\theta\,\rho(\theta)=1$, the entropy density would have the exact root-density representation
\begin{equation}
s_2=\log2-\frac12\int d\theta\,\rho(\theta)
\log\!\left(4\sin^2\frac{\theta}{2}\right).
\label{eqS:rootdensityentropy}
\end{equation}
The representation is exact but not presently an evaluation, because the full limiting density $\rho(\theta)$ is not known in closed form; this is why the convergent Ising/cluster route is more effective.

\subsection{Why the squared minors do not collapse to one ordinary determinant}

A natural hope would be a matrix $B$ satisfying
$\det B[S,S]=d_L(S)^2$ for every $S$, which would make $Q_L(u)=\det(I+uB)$.  Already two-site minors exclude a real symmetric realization.  One-site minors require $B_{ii}=A_L^2$, while
\begin{equation}
d_L(\{i,j\})^2=A_L^4[g_{i-j}^{(L)}]^2>A_L^4.
\end{equation}
A real symmetric $B$ would instead give
$\det\bigl(\begin{smallmatrix}A_L^2&B_{ij}\\B_{ij}&A_L^2\end{smallmatrix}\bigr)
=A_L^4-B_{ij}^2\le A_L^4$.  The obstruction is local: the polynomial simplifications above are not disguised ordinary determinantal point-process identities.

\section{Connections to Selberg sums, free-fermion counting, and related exact structures}

The root-of-unity form of the finite Cauchy matrix places the all-minors problem close to discrete Selberg/Dyson sums \cite{Selberg1944,deBruijn1955,ForresterWarnaar2008}.  The present route is complementary: rather than evaluating selected finite-size Selberg sums, it turns the probability problem into a weak classical Ising/current model, giving exact finite-size access and explicit thermodynamic pieces at arbitrary real $n>1$.

The probability determinants also sit naturally beside emptiness formation probabilities and full counting statistics in free fermions \cite{AbanovFranchini2003,FranchiniAbanov2005,StephanEFP2014,Ivanov2013,Groha2018,AresViti2020}.  Individual Gaussian probabilities are polynomial-time objects \cite{NajafiRajabpour2016,TarighiKhassehRajabpour2024,NattaghNajafi2025}.  The difficulty addressed here is the nonlinear sum over all probabilities.  Participation and multifractality in quantum spin chains provide the broader many-body context \cite{AtasBogomolny2012,AtasBogomolny2014,Misguich2016,LuitzParticipation2014,AlcarazRajabpour2015,Alcaraz2016,MisguichOshikawa2017,SierantTurkeshi2022}, while earlier information-theoretic studies of measured spin configurations and classical critical Ising systems include Refs.~\cite{UmParkHinrichsen2012,LauGrassberger2013}.  Real-space RG has also been used to study the leading Shannon--R\'enyi term of the pure and random quantum Ising chain \cite{Monthus2015}.

\section{Finite-size checks}

We performed five independent algebraic checks before taking asymptotic limits.

First, for several integer $n$ and finite $L$, direct polynomial evaluation of Eq.~\eqref{eqS:Bnfull} agrees with the root product Eq.~\eqref{eqS:Bfactor} to machine precision.  Second, for several values of $\lambda$ the finite product Eq.~\eqref{eqS:finiteSine} agrees with direct multiplication over all $m=1,\ldots,L-1$.  Third, direct Walsh enumeration from Eq.~\eqref{eqS:Walsh}, using the exact Cauchy products rather than diagonalizing the Hamiltonian, gives normalized probabilities with exactly half of the configurations supported, as required by the even global parity sector.  Fourth, an independent enumeration of Bernoulli edge sets in Eq.~\eqref{eqS:paritylaw} reproduces every determinant probability at $L=4$ and $L=6$ to machine precision.  Finally, the single-cycle formula Eq.~\eqref{eqS:singlecycleReal} agrees with direct parity-fiber summation for noninteger as well as integer $n$.

Representative entropies obtained by this independent Walsh enumeration at $L=16$ are
\begin{center}
\begin{tabular}{c c}
\toprule
$n$ & $H_n^{\rm PBC}(16)$\\
\midrule
$1.2$ & $5.44036717785$\\
$1.5$ & $4.39772916797$\\
$2$ & $3.39941011187$\\
$3$ & $2.62473721490$\\
$4$ & $2.33613398969$\\
\bottomrule
\end{tabular}
\end{center}
For $n=2,3,4$, size sequences are consistent with a vanishing intercept after the expected $1/L$ corrections.  For $1<n<3/2$, the visibly slower convergence is consistent with the $L^{2-2n}$ correction in Eq.~\eqref{eqS:branchcorrection}; this is precisely why fitting those indices with only integer powers of $1/L$ can create a spurious apparent constant.

To probe specifically the residual real-index sector, we also evaluate the exact ratio $\Xi_L(n)$ of Eq.~\eqref{eqS:XiRealDef} with the same Walsh enumeration.  The size-difference estimator
\begin{equation}
\psi_n^{\rm eff}(L)=\frac12\left[\log\Xi_L(n)-\log\Xi_{L-2}(n)\right]
\label{eqS:psieffreal}
\end{equation}
approaches the residual bulk pressure if that thermodynamic limit is regular.  Representative values are
\begin{center}
\begin{tabular}{c c c c}
\toprule
$n$ & $\psi_n^{\rm eff}(12)$ & $\psi_n^{\rm eff}(16)$ & $\psi_n^{\rm eff}(20)$\\
\midrule
$1.10$ & $0.005490$ & $0.005348$ & $0.005273$\\
$1.20$ & $0.006347$ & $0.006082$ & $0.005956$\\
$1.30$ & $0.005482$ & $0.005222$ & $0.005119$\\
$1.40$ & $0.004231$ & $0.004058$ & $0.004014$\\
$1.50$ & $0.003106$ & $0.003040$ & $0.003052$\\
$1.75$ & $0.001379$ & $0.001478$ & $0.001551$\\
$2.00$ & $0.000688$ & $0.000809$ & $0.000873$\\
\bottomrule
\end{tabular}
\end{center}
Across the noninteger range $1.1\le n<2$, the residual-sector estimates remain small and vary smoothly with both $n$ and $L$; no additional rapidly growing finite-size contribution is visible through $L=20$.  These data are supporting evidence only and are not used as a substitute for the all-orders convergence question discussed above.

\section{Kramers--Wannier dual basis}

The exact Kramers--Wannier relation for the TFI Shannon--R\'enyi entropies \cite{Stephan2010} is
\begin{equation}
H_n^{(x)}(h)=H_n^{(z)}(1/h)+\log2.
\label{eqS:KW}
\end{equation}
At the self-dual point $h=1$, the $n=2$ result therefore gives
\begin{equation}
c_2^{\mathrm{PBC},x}=\log2,
\qquad c_2^{\mathrm{PBC},z}=0.
\label{eqS:KWconstants}
\end{equation}
For the broader $n>1$ branch, the previously observed $x$-basis constant $\log2$ is equivalently the statement that the continued $z$-basis constant is zero \cite{Stephan2010}.  We use this as an external consistency check on the continuation, not as a substitute for the lattice proof at $n=2$.

\end{document}